\documentclass[conference]{IEEEtran}
\IEEEoverridecommandlockouts
\usepackage[numbers,sort&compress]{natbib}
\usepackage{booktabs}
\usepackage{graphicx}
\usepackage{multirow}
\usepackage{amsmath,amssymb,amsfonts}
\usepackage{algorithmic}
\usepackage{graphicx}
\usepackage{bm}
\usepackage{xcolor,colortbl}
\usepackage{textcomp}
\usepackage[export]{adjustbox}
\usepackage{xcolor}
\usepackage[acronym]{glossaries}    

\newacronym {mwn}       {MWN}       {Mobile Wireless Network}

\newacronym {lte}       {LTE}       {Long-Term Evolution}
\newacronym {nr}        {NR}       {New Radio}
\newacronym {1g}        {1G}        {1$^{st}$ Generation}
\newacronym {2g}        {2G}        {2$^{nd}$ Generation}
\newacronym {3g}        {3G}        {3$^{rd}$ Generation}
\newacronym {4g}        {4G}        {4$^{th}$ Generation}
\newacronym {5g}        {5G}        {5$^{th}$ Generation}
\newacronym {b5g}       {B5G}       {Beyond 5$^{th}$ Generation}
\newacronym {6g}        {6G}        {6$^{th}$ Generation}

\newacronym {3gpp} 		{3GPP} 		{3$^{rd}$ Generation Partnership Project}
\newacronym {r14}       {R14}       {Release 14}
\newacronym {r15}       {R15}       {Release 15}
\newacronym {r16}       {R16}       {Release 16}
\newacronym {r17}       {R17}       {Release 17}
\newacronym {r18}       {R18}       {Release 18}
\newacronym {r19}       {R19}       {Release 19}

\newacronym{itu}        {ITU}       {International Telecommunication Union}

\newacronym{genai}     {GAI}     {Generative Artificial Intelligence}
\newacronym{ai}        {AI}      {Artificial Intelligence}
\newacronym{qa}        {QA}      {Question-Answering}
\newacronym{llm}       {LLM}     {Large Language Model}
\newacronym{slm}       {SLM}     {Small Language Model}
\newacronym{rag}       {RAG}     {Retriever-Augmented Generation}
\newacronym{nlp}       {NLP}     {Natural Language Processing}
\newacronym{bert}      {BERT}    {Bidirectional Encoder Representations from Transformers}
\newacronym{sbert}     {SBERT}      {Sentence Bidirectional Encoder Representations from Transformers}
\newacronym{roberta}   {RoBERTa}    {Robustly Optimized BERT Approach}
\newacronym{gpt}       {GPT}    {Generative Pre-trained Transformer}
\newacronym{ir}        {IR}     {Information Retrieval}
\newacronym{tfidf}     {TF-IDF} {Term Frequency-Inverse Document Frequency}
\newacronym{bm25}      {BM25}   {Best Matching 25}
\newacronym{dsft}      {DSFT}   {Domain-Specific Fine-Tuning}
\newacronym{dpr}       {DPR}    {Dense Passage Retriever}
\newacronym{dhr}       {DHR}    {Dense Hierarchical Retrieval}
\newacronym{dhrd}      {DHR-D}  {Dense Document-level Retrieval}
\newacronym{dhrp}      {DHR-P}  {Dense Passage-level Retrieval}
\newacronym{mrr}       {MRR}    {Mean Reciprocal Rank}
\newacronym{ts}        {TS}     {Technical Specification}

\newacronym {cdf} 	   {CDF}    {Cumulative Density Function}

\newacronym {mcq} 	   {MCQ}    {Multiple Choice Question}
\newacronym {json} 	   {JSON}   {JavaScript Object Notation}
\newacronym {api}      {API}    {Application Programming Interface}

\def\BibTeX{{\rm B\kern-.05em{\sc i\kern-.025em b}\kern-.08em
    T\kern-.1667em\lower.7ex\hbox{E}\kern-.125emX}}

\begin{document}

\title{Telco-DPR: A Hybrid Dataset for Evaluating Retrieval Models of 3GPP Technical Specifications}


\author{\IEEEauthorblockN{T. Saraiva$^{1,2,4}$, M. Sousa$^{2,3}$}
\IEEEauthorblockA{$^{1}$\textit{Instituto de Telecomunica\c{c}\~{o}es} \\
$^{2}$\textit{CELFINET - A Cyient Company}\\
Lisbon, Portugal \\
{[thaina.saraiva, marco.sousa]@cyient.com}}
\and
\IEEEauthorblockN{P. Vieira$^{1,3}$}
\IEEEauthorblockA{$^{3}$\textit{Instituto Superior de Engenharia Lisboa}\\
Lisbon, Portugal\\
pedro.vieira@isel.pt}
 
\and
\IEEEauthorblockN{A. Rodrigues$^{1,4}$}
\IEEEauthorblockA{
$^{4}$\textit{Instituto Superior T\'{e}cnico}\\
Lisbon, Portugal\\
ar@lx.it.pt}
}

\maketitle

\begin{abstract}
This paper proposes a \gls{qa} system for the telecom domain using \gls{3gpp} technical documents. Alongside, a hybrid dataset, \textit{Telco-DPR}\footnote{\texttt{Telco-DPR} dataset presented in this paper is available open-source \cite{sourceDAtaset}.}, which consists of a curated \gls{3gpp} corpus in a hybrid format, combining text and tables, is presented. Additionally, the dataset includes a set of synthetic question/answer pairs designed to evaluate the retrieval performance of \gls{qa} systems on this type of data. The retrieval models, including the sparse model, \gls{bm25}, as well as dense models, such as \gls{dpr} and \gls{dhr}, are evaluated and compared using top-K accuracy and \gls{mrr}. The results show that \gls{dhr}, a retriever model utilising hierarchical passage selection through fine-tuning at both the document and passage levels, outperforms traditional methods in retrieving relevant technical information, achieving a Top-10 accuracy of 86.2\%. Additionally, the \gls{rag} technique, used in the proposed \gls{qa} system, is evaluated to demonstrate the benefits of using the hybrid dataset and the \gls{dhr}. The proposed \gls{qa} system, using the developed \gls{rag} model and the \gls{gpt}-4, achieves a 14\% improvement in answer accuracy, when compared to a previous benchmark on the same dataset. 
\end{abstract}

\glsresetall


\section{Introduction}
\label{sec:intro}

As research efforts towards \gls{6g} networks intensify, \gls{ai} and \gls{genai} are emerging as unique enablers. They promise to enhance essential network functions such as planning, operation, and optimisation, while facilitating the development of new services and use cases for \gls{6g} and also for legacy networks~\cite{bariah2023large, maatouk2024largelanguagemodelstelecom, tarkoma2023ainativeinterconnectframeworkintegration, zhou2024large}. Meanwhile, the \gls{3gpp}, which has overseen the development of universal standards for \glspl{mwn}, such as the current \gls{5g}, continues to lead the standardisation of current and future generations of \glspl{mwn}~\cite{3gpp_specifications}.

A possible \gls{genai} application for the operation and optimisation of \glspl{mwn} is on \gls{qa} systems, which are designed to respond to advanced user's questions about \glspl{mwn} automatically, using natural language. Powered by \glspl{llm}, these systems have proven highly successful in open-domain contexts. Yet, they face difficulties when applied to domain-specific contexts such as \glspl{mwn}. \glspl{llm}, typically trained on general-purpose corpora, struggle with the specialised language, technical jargon, and complex protocols that define technical standards like those established by the \gls{3gpp}~\cite{kandpal2023largelanguagemodelsstruggle}. This limitation entails a significant challenge in developing \glspl{qa} systems that accurately access and understand the specialised information of \glspl{mwn}.

\gls{qa} systems for domain-specific contexts can be developed by fine-tuning \glspl{llm} with domain specific data sources~\cite{roberts2020knowledgepackparameterslanguage}. However, this method is computationally intensive and might be insufficient for highly specialised tasks. A more efficient alternative is using approaches based on \gls{rag} ~\cite{dai2019transformerxlattentivelanguagemodels}. Firstly, a retriever system that identifies and retrieves relevant passages from a knowledge base to assist in answering a question is employed. Then, the retrieved passages go to a reader system (often a transformer-based model) which leverages the retrieved contextual information to interpret and generate the appropriate answer to the original question. While \gls{rag} effectively handles unstructured text, most technical documents, as in \glspl{mwn}, also include structured elements like tables. These tables are typically interspersed with the text and are cross-referenced within the same document, providing essential supplementary or complementary semantic information~\cite{chen2021openquestionansweringtables}.

This paper presents a \gls{qa} system specifically tailored to the \gls{3gpp} standards by leveraging \gls{rag}, while addressing the structured and hybrid nature of telecommunication (telecom) standardisation documents. By incorporating tables as essential components in the \gls{qa} process, this system seeks to overcome the limitations of naive-\gls{rag} frameworks, and enhance the retrieval and comprehension of highly technical information in the field of \glspl{mwn}. 
Thus, the main contributions of this paper are:
\begin{itemize} 
    \item Presentation of Telco-DPR~\cite{sourceDAtaset}, a public dataset containing a curated hybrid corpus of \gls{3gpp} documents, in tandem with a set of synthetic question-answer pairs to enable the evaluation of retrieval models.
    \item Proposal of a \gls{qa} system specifically tailored to \gls{3gpp} standards, leveraging the \gls{rag} framework with a hybrid corpus and a hierarchical retriever.

\end{itemize}

The paper is organised as follows. After the introduction, Section~\ref{sec:sota} briefly presents the fundamental concepts involving the proposed methodology for \gls{qa} systems and its related work. Section~\ref{sec:methodology} presents the proposed methodology. Section~\ref{sec:results} presents the main results, while Section~\ref{sec:conclusions} concludes the paper and highlights future work. 
\section{Fundamentals and Related Work}
\label{sec:sota}
This section presents the fundamental concepts involved in the proposed methodology for the \gls{qa} system and a brief literature review of related work.

\subsection{Question-Answering Systems}
Early \gls{qa} systems employed traditional methods, such as rule-based approaches~\cite{riloff2000rule}. However, recent \gls{qa} systems have shifted towards \gls{rag} techniques based on a general retrieve-then-read architecture \cite{zhu2021retrievingreadingcomprehensivesurvey}, where a retriever selects relevant documents or passages from a knowledge base, and a reader extracts or generates answers based on the provided context and question. The concepts of passages, tokens, and the main two elements of the retrieve-then-read architecture are defined next.

\subsubsection{Passages and Tokens}
In \gls{qa} systems using \gls{rag}, passages are the fundamental units, commonly referring to a meaningful text segment (\textit{e.g.,} paragraph) containing sufficient context for answering a user question. These passages result from processing a knowledge base such as \gls{3gpp} \glspl{ts} and segmenting all the information into passages. Along with passages, tokens are paramount to \gls{qa} systems that leverage \glspl{llm}. Tokenization breaks down text into units (tokens), which can be words or sub-words, depending on the employed method. Ensuring that the tokenized versions of passages fit within the token limit of the used \glspl{llm} is crucial; passages that are too long, may be truncated, while overly short passages might miss relevant information.

\subsubsection{Retriever} The primary aim of a retriever is to identify the relevant passages from large knowledge bases. Several retrieval techniques have been developed, from traditional \gls{ir} methods to modern deep learning-based approaches. Classic \gls{ir} models, such as \gls{tfidf} and \gls{bm25}~\citep{harman2019information, robertson2009probabilistic, sciavolino2022simpleentitycentricquestionschallenge} rank results based on relevance metrics and the frequency of question terms. These sparse methods perform well in keyword-based retrieval, but tend to struggle with the semantic meaning of natural language questions, limiting their effectiveness.

In contrast to sparse retrieval, \gls{dpr} leverages deep learning to enhance comprehension of semantic meaning, enabling the generation of dense vector embeddings for both questions and passages, employing models such as \gls{sbert}~\cite{tonellotto2022lecture}. These embeddings map questions and passages into a shared latent space, allowing the retriever to find semantically relevant passages by comparing vector similarity, even in cases where the terms in the query and passage do not directly overlap.



Advanced retrievers built upon \gls{dpr}, such as \gls{dhr} \cite{liu2021densehierarchicalretrievalopendomain}, improve the passage selection process in documents with a well-defined structure. The \gls{dhr} utilizes a two-stage hierarchical approach: in the first stage, \gls{dhrd} assesses the similarity between the question and the document's metadata. In the second stage, \gls{dhrp} focuses the search on the passages within the most relevant documents. Thus, \gls{dhr} offers both a macroscopic and microscopic understanding of semantics.

\subsubsection{Reader} The task of a reader is to generate an answer based on the context provided by the retriever, \textit{i.e.}, the retrieved passages, and the user question. There are two main types of readers: extractive and generative. Extractive readers are trained to extract the correct answer directly from the retrieved passage. For a question $q$ and context $c$ consisting of $n$ tokens, the reader identifies the start and end tokens that form the answer~\cite{baradaran2020surveymachinereadingcomprehension}. Extractive readers excel in scenarios where the answer is explicitly contained in the passage.

Generative readers, leveraging \glspl{llm} such as \gls{bert} and \gls{gpt}, are widely used with RAG \cite{lewis2021retrievalaugmentedgenerationknowledgeintensivenlp}. They combine the questions and retrieved passages as input and generate answers in natural language. In this approach, the top retrieved passages are concatenated with the question and fed into the \gls{llm} to produce an answer. This makes generative readers particularly suitable for scenarios where answers need to be inferred or rephrased, rather than directly extracted.  


\subsection{Related Work}
The first public telecom \gls{qa} dataset, TeleQnA, was introduced in~\cite{maatouk2023teleqna}, featuring \glspl{mcq} derived from research articles and technical documents, such as \gls{3gpp} \glspl{ts}.  Enhancements involving query augmentation and indexing strategies were proposed in~\cite{bornea2024telcoragnavigatingchallengesretrievalaugmented}, resulting in an accuracy improvement of up to 14.45\% on a subset of the \glspl{mcq} from the TeleQnA dataset, compared to the original work in~\cite{maatouk2023teleqna}. The joint usage of \glspl{slm} and \gls{rag} was explored in~\cite{piovesan2024telecomlanguagemodelslarge}, showing that these models can achieve accuracy comparable to that of \glspl{llm}.

In~\cite{nikbakht2024tspecllmopensourcedatasetllm}, the authors demonstrated that even a naive-\gls{rag} applied to \gls{qa} improved response accuracy from 46\% to 75\% using Gemini 1.0 on their synthetic QA dataset, known as Q-Small. Similarly, the authors in ~\cite{yilma2024telecomragtamingtelecomstandards} presented a \gls{rag}-based \gls{qa} system for \gls{3gpp} \glspl{ts}, achieving improved accuracy. However, this study focused primarily on qualitative results. Finally, the authors in~\cite{doringer2023synthetic} compared retrievers used in \gls{rag}, specifically dense retrievers based on \gls{bert} and \gls{bm25}-based sparse retrievers, demonstrating significant improvements in Top-K retrieval accuracy, when considering the first ten retrieved passages ($K = 10$). However, the dataset used in this study remains private, which limits broader evaluation.

Regarding \gls{3gpp} text corpora, two publicly available datasets exist: SPEC5G and TSpec-LLM. 
The SPEC5G is a dataset extracted from 3GPP \glspl{ts} presented in~\cite{karim2023spec5gdataset5gcellular}. While primarily used for summarising and classification, SPEC5G can also be employed as a text corpus for \gls{qa} systems. However, the corpus is presented as a single text file, lacking the granularity of individual \glspl{ts}. This limitation hampers the effectiveness of text splitting in \gls{rag}-based systems. Further advancements are detailed in~\cite{nikbakht2024tspecllmopensourcedatasetllm}, where a text corpus dataset incorporating document source metadata was created in \texttt{Markdown} format from the \glspl{ts} of multiple \gls{3gpp} releases. However, this format prevents direct separation of text and table elements.

\section{Methodology}
\label{sec:methodology}
This section presents the methodology for the proposed \gls{qa} system applied to \glspl{mwn}, as depicted in Fig.~\ref{fig:flowchart}. The methodology begins with the data extraction from \gls{3gpp} \glspl{ts} to create a hybrid \gls{3gpp} corpus (knowledge base), followed by the generation of pairs of questions and answers. The corpus and the question/answer pairs constitute a dataset for implementing and optimising \gls{qa} systems, which is now publicly available on the \textit{Hugging Face} repository \cite{sourceDAtaset}. Next, the base architecture of the \gls{qa} system is defined and optimised using the corpus and the question/answer pairs, leading to the final \gls{qa} system for \glspl{mwn}. Its performance is evaluated on a public \gls{qa} dataset.

\subsection{3GPP Corpus}
Transforming technical documents into machine-readable formats is a crucial step for enabling a \gls{qa} system to efficiently search for information and deliver accurate answers. A key characteristic of the \gls{3gpp} \glspl{ts} is that each release contains several documents in \texttt{.docx} format, with a distinct scope and a well-defined structure, as indicated by a table of contents at the beginning. Additionally, these documents are heavily populated with tables, equations, and figures.

The \gls{3gpp} \glspl{ts} were extracted from the source site~\cite{download_3gpp} using an automated extraction tool. Then, a Python script, leveraging the \texttt{python-docx} library~\cite{python_docx}, was used, enabling the tagging of various elements such as paragraphs, lists, tables, and images. During this extraction phase, several challenges were encountered, particularly with equations and figures. The text embedded within equations could not be retrieved as plain text, and the resulting images had low quality. Consequently, the equations and images were excluded from the conversion process, and only text-based elements were retained for further processing.

The tagging elements produced by the extraction tool enabled the differentiation between text passages and table elements. The paragraph tag was utilised to create text passages, with a token limit of 512 tokens established for each passage. In instances of longer paragraphs, the semantic similarity-based splitting mechanism described in \cite{agirre2009study} was employed, while shorter paragraphs within the same section were aggregated (\textit{e.g.,} small list items).
To preserve the structural integrity of the tables, a multi-vector retriever for tables was employed~\cite{langChain}, indexing each table using both its caption and a \gls{llm}-generated summary of its contents. This approach establishes the semantic relationship between the two types of data \cite{min2024exploringimpacttabletotextmethods}.

\begin{figure}[!t]
    \centering
    \includegraphics[width=0.63\linewidth]{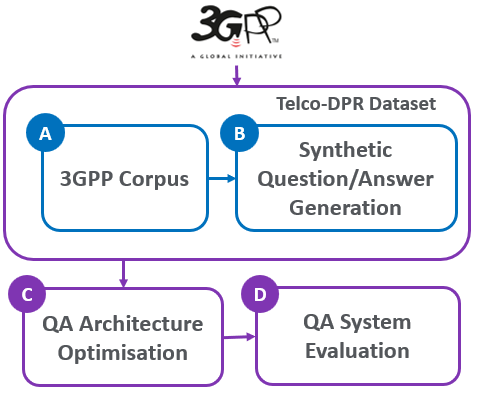}
    \caption{Development methodology for the proposed QA system.}
    \label{fig:flowchart}
\end{figure}

As shown in Fig.~\ref{fig:flowchart}, this process results in a curated hybrid dataset, called Telco-DPR. 
The passages in this dataset are enriched with relevant metadata, including the document source, release number, and section/subsection title. For passages that serve as captions for tables, an index is provided to identify the table's location within the dataset.

\subsection{Synthetic Question/Answer Generation}
\label{subsec:qadataset}

A subset of four \glspl{ts} from the hybrid \gls{3gpp} corpus was used in this step. Passages from these \glspl{ts} were provided to a \gls{llm}, and a specifically designed prompt was used to generate between one and five questions from each passage, creating a set of question-answer pairs. Notably, each question/answer pair includes an identifier for the corresponding passage used to create the question, which is the base for evaluating the performance of retriever models used in RAG. This process produced both many-to-one and one-to-one mappings between questions and passages. 

The prompt was carefully crafted to guide the \gls{gpt}-3.5 in generating diverse and accurate questions. For passages containing table captions, the original tables were included in \texttt{Markdown tables} format to ensure that the generated questions accurately reflected the tabular content. After generating the question/answer pairs, a post-processing step was conducted to filter out any irrelevant or incorrect questions produced by the \gls{llm}, with validation performed by human domain experts.


\subsection{QA Architecture Optimisation}
The proposed architecture for the \gls{qa} system adopts the widely used retriever-then-reader framework. However, due to the specific nature of \gls{3gpp} \glspl{ts}, an optimisation process was essential to refine the system’s architecture to answer technical user questions straightforward. Regarding the reader component, a generative reader was chosen for its ability to handle the semantic complexity required for technical question answering, leveraging the capabilities of \glspl{llm}. Therefore, the primary focus of the optimisation was to improve the architecture retriever component. To this end, several retriever methods were evaluated based on their ability to retrieve the correct passages from the \gls{3gpp} hybrid corpus using the question/answer pairs from the Telco-DPR dataset (see Section~\ref{subsec:qadataset}). 

The first considered retriever method was the \gls{bm25}, which served as the baseline for comparison with more advanced methods. Next, a dense retrieval method, \gls{dpr}, was implemented using a pre-trained transformer-based embedding model, \gls{sbert}, specifically ``\textit{multi-qa-MiniLM-L6-cos-v1}'' \cite{sbertModel}. SBERT has proven to be an effective alternative to LLM-based word embeddings, offering efficiency when computational resources are limited \cite{freestone2024wordembeddingsrevisitedllms}.

A key advantage of dense retrieval methods is that the embedding models, based on neural networks, can be fine-tuned to domain-specific data. Thus, two versions of the \gls{dpr}, with and without fine-tuning of the \gls{sbert}, were implemented. In the fine-tuned process, the ``\textit{Multiple Negatives Ranking Loss}'' was used as the loss function \cite{henderson2017efficientnaturallanguageresponse}. This loss function is optimised by contrasting an anchor passage with both positive and negative passages. the goal is to minimise the distance between the anchor (the question) and the positive passage (which contains the answer), while maximising the distance between the anchor and the negative passage (which is similar but does not contain the answer).
%

Lastly, the \gls{dhr} approach was considered at both the document and passage levels, \gls{dhrd} and \gls{dhrp}, respectively. For the document-level retriever, the data representation consisted of the concatenation of the document title, document abstract, and section/subsections titles, separated by a reserved token. For the passage-level retriever, the data representation was formed by concatenating the section/subsections titles and the passage content. The \gls{sbert} embedding model \cite{sbertModel} was also employed in this approach. 

The performance of all retriever models was evaluated using the Top-$K$ accuracy and the \gls{mrr} ~\cite{harman2011information}. The top-$K$ accuracy measures the proportion of times the correct passage appears within the first $K$ passages retrieved. The \gls{mrr} is  the average of the inverse of the rank position, evaluates the rank of the first correct passage, ranging from 0 (worst) to 1 (best). These metrics provide the basis for comparing the considered retrieval models.



\subsection{QA System Evaluation}
The proposed \gls{qa} system was evaluated using the public \gls{qa} questionnaire, called Q-Small dataset, introduced in~\cite{nikbakht2024tspecllmopensourcedatasetllm}. This questionnaire was chosen due to its alignment with the content of the Telco-DPR, which is grounded in \gls{3gpp} \gls{r15} and \gls{r16}, and consists of 100 \glspl{mcq}. The questions are divided into three categories based on their difficulty: easy (30 questions), intermediate (51 questions), and hard (19 questions).
Another important characteristic of this questionnaire is that 74\% of the answers are found within tables, requiring reader models to understand the table structure in order to accurately provide the correct answers.


The performance of the proposed \gls{qa} system, which employs \gls{rag} with hybrid data, was evaluated based on the accuracy of the generated answers. These results were then compared to the benchmark, established in \cite{nikbakht2024tspecllmopensourcedatasetllm}.
 
\section{Results}
\label{sec:results}
This section presents the results of the proposed methodology, relative to: the generation of the Telco-DPR dataset, the optimisation of the \gls{qa} system for \glspl{mwn} (focused on the retriever models), and the end-to-end \gls{qa} system evaluation.

\subsection{Telco-DPR dataset}
The Telco-DPR dataset comprises a corpus of 57 3GPP \glspl{ts}, where 36 belong to \gls{r15}, and 21 are from \gls{r16}, totalling 14.654 passages. The corpus contains 2.316 tables, averaging 40.63 tables per document. Each table contains approximately 850 tokens. In contrast, textual passages, average 177 tokens.

The Telco-DPR synthetic \gls{qa} dataset pairs each question with the relevant passage. The \gls{api} of \gls{gpt}-3.5 was used to generate 1.741 question-answer pairs based on four 3GPP \glspl{ts}. The dataset was split into 1.218 examples for fine-tuning and 523 for testing. Table \ref{tab:statisticQuestions} provides statistics on the questions and passages, including the distribution of questions generated across the four \glspl{ts}, amount of tokens and the frequency of interrogative words. The majority of questions begin with ``What'' or ``Which'', indicating a focus on factual information. ``How'' questions, while less common, often seek procedural or methodological details. The average number of tokens for questions and answer is 21 and 147 tokens, respectively.

\begin{table}[!t]
\caption{Statistics of Questions in Telco-DPR.}
\label{tab:statisticQuestions}
\centering
\renewcommand{\arraystretch}{1.1}  
\setlength{\tabcolsep}{10pt}        
\begin{tabular}{ll|cc}
\hline
\multicolumn{2}{c|}{\textbf{Question by TS (\%)}} & \multicolumn{2}{c}{\textbf{First Question Word (\%)}} \\ \hline
TS 36.777 (R15) & 38  & What & 79.67 \\
TS 38.811 (R15) & 26  & Which & 9.60 \\
TS 38.821 (R16) & 17  & How & 8.61 \\
TS 38.901 (R16) & 19  & Is/Are & 0.63 \\ \cline{1-2}
\multicolumn{2}{c|}{\textbf{Average Tokens}} & Where & 0.40 \\ \cline{1-2}
Question & 21  & Why & 0.34 \\
Answer & 147  & Others & 0.75 \\ \hline
\end{tabular}
\end{table}

\subsection{QA Architecture Optimisation}
The optimization of the \gls{qa} architecture considered several retriever models: \gls{bm25}, \gls{dpr}, \gls{dpr} Fine-Tuned (\gls{dpr}-FT), \gls{dhrp}, and \gls{dhr} (incorporating both \gls{dhrd} and \gls{dhrp}). These models were evaluated on the test subset of the Telco-DPR using as reference the Top-$K$ accuracy and the \gls{mrr} to measure their ability in retrieving the correct passages for the corresponding questions.  

Table \ref{tab:results} presents the evaluation metrics for each retriever model. \gls{bm25} shows relatively lower accuracy, achieving 26.3\% at Top-1 (Acc@1) and 59.2\% at Top-10 (Acc@10), with \gls{mrr} at Top-10 (MRR@10) of 0.36. This is mainly due to the reliance on exact word matching rather than understanding the context. The \gls{dpr} has a higher performance, yielding a MRR@10 of 0.45. \gls{dhrp} leads to further improvements by enriching the passage representation through concatenation of the section title with the text passage. The \gls{dhr} (\gls{dhr}-D and \gls{dhr}-P), demonstrate the best performance across all metrics, achieving an Acc@10 of 86.2\% and the highest MRR@10 of 0.68. The \gls{dhr} retrieval demonstrates, the highest performance in retrieving the correct passages for answering user questions. Thus, the proposed QA system leverages this retrieval model.

\begin{figure}[!ht]
\centering
    \includegraphics[width=0.7\linewidth]{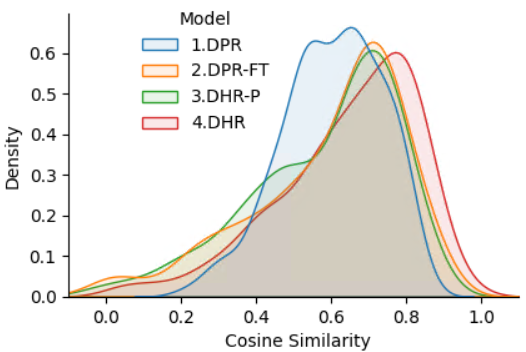}
    \caption{The distribution of cosine similarities between
question and answer for dense retriever models.}
\label{fig:similarity}
\end{figure}

Additionally, to analyse the performance differences among the four dense retrieval models, Fig. \ref{fig:similarity} presents a comparison of cosine similarity distributions, metric used in similarity search to select the most relevant passages, between the question and the passage containing the answer. The DPR model (blue) shows a broad distribution of lower similarity scores, while the fine-tuned models shift towards higher scores. The DHR model (red) stands out with a high concentration of scores between 0.7 and 0.9, indicating improvement by increase the similarity scores for the most relevant passages.


\subsection{QA System Evaluation}
\begin{figure}[!t]
    \centering
    \includegraphics[width=0.85\linewidth]{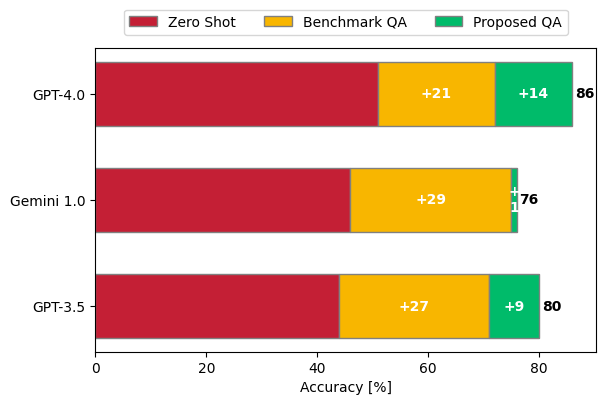}
    \caption{A comparison of the accuracy of the proposed QA system, the benchmark \gls{qa} against the zero-shot of \gls{gpt}-3.5, Gemini 1.0, and \gls{gpt}-4.0.}
    \label{fig:rag-results}
\end{figure}

Fig. \ref{fig:rag-results} presents the accuracy of the proposed \gls{qa} system (using Top-10 passages retrieved by \gls{dhr} as context), alongside the benchmark QA system proposed in \cite{nikbakht2024tspecllmopensourcedatasetllm} and three \glspl{llm} operating in a zero-shot scenario, relying solely on their pre-existing knowledge: \gls{gpt}-3.5, Gemini 1.0, and \gls{gpt}-4. 
While \gls{gpt}-4 excelled in zero-shot learning, Gemini 1.0 demonstrated superior accuracy within the benchmark \gls{qa}. The proposed QA system, utilising \gls{gpt}-4, achieved  86\% accuracy, surpassing the benchmark \gls{qa} by 14\%. This result highlights \gls{gpt}-4's ability to handle effectively hybrid data.

To evaluate the accuracy of the retrieved passages, the correct passage associated with each question was identified within the Telco-DPR dataset. Table \ref{tab:retriever_metrics_by_level} presents the accuracy rankings and \gls{mrr} for the question-answer retrieval task. The results show that 78\% of the questions had the correct passages retrieved within the Top-10, with an MRR@10 of 0.56. Interestingly, the ``Easy'' questions, despite their lower complexity, exhibited lower accuracy, whereas the ``Hard'' questions achieved the highest accuracy at 94.7\%. 
Regarding the final accuracy of the QA system using RAG with GPT-4, the performance was consistent across different question difficulty levels. The highest accuracy was 87\% for easy questions, while the lowest was 84\% for hard questions.

A deeper analysis reveals that the QA system responses were not always grounded in the correct context. As shown in Table \ref{tab:retriever_metrics_by_level}, the maximum accuracy at rank 10 is 78\% (78 questions). Notably, the \gls{rag} with \gls{gpt}-4, achieving an accuracy of 86\% (86 questions), successfully answered 70 questions based on the correct context, failing to infer 8.


\begin{table}[!t]
\caption{Retriever Performance on the Telco-DPR Dataset.}
\label{tab:results}
\centering
\renewcommand{\arraystretch}{1.2}  
\setlength{\tabcolsep}{6pt}        
\begin{tabular}{lccccc}
\hline
\textbf{Retriever} & \textbf{Acc@1}& \textbf{Acc@3} & \textbf{Acc@5} & \textbf{Acc@10} & \textbf{MRR@10} \\ \hline
BM25              & 26.3        & 43.0        & 49.9        & 59.2         & 0.36         \\ 
DPR               & 35.5        & 51.0        & 58.8        & 67.3         & 0.45        \\ 
DPR-FT            & 58.8        & 69.9        & 75.9        & 80.5         & 0.65         \\ 
DHR-P             & 57.9        & 71.8        & 77.0        & 82.0         & 0.66         \\ 
\textbf{DHR}      & \textbf{59.2} & \textbf{75.1} & \textbf{81.8} & \textbf{86.2} & \textbf{0.68} \\ \hline
\end{tabular}
\end{table}

\begin{table}[!ht]
\caption{Evaluation of Retriever Models and QA System with Hybrid Data and DHR.}
\label{tab:retriever_metrics_by_level}
\centering
\renewcommand{\arraystretch}{1.5} 
\setlength{\tabcolsep}{6pt}       
\begin{tabular}{lcccl}
\hline
\multicolumn{5}{c}{\textbf{Retriever Performance at Rank 10}}                                 \\ \hline
\textbf{}          & \textbf{Easy} & \textbf{Intermediate} & \textbf{Hard} & \textbf{Overall} \\ \hline
\textbf{Accuracy}  & 66.6\%        & 78.4\%                & 94.7\%        & 78.0\%           \\
\textbf{MRR}       & 0.43          & 0.61                  & 0.65          & 0.56             \\ \hline
\multicolumn{5}{c}{\textbf{QA System Accuracy}}                                               \\ \hline
\textbf{RAG+GPT-4} & 87.0\%        & 86.0\%                & 84.0\%        & 86.0\%           \\ \hline
\end{tabular}

\end{table}

Furthermore, 16 of the correctly answered questions were inferred without relying on the correct context, suggesting potential reliance on the LLM own memory or the generation of random responses. Consequently, the \gls{mcq} dataset may have limitations in accurately assessing the true quality of \gls{rag} systems. A more effective evaluation could be achieved using open-ended questions for evaluation. However, the scarcity of reliable telecom-specific datasets presents an opportunity for continuous research in this area.


\section{Conclusion}
\label{sec:conclusions}
This paper presents a novel \gls{qa} system tailored to answer questions related to \gls{3gpp} \glspl{ts}. Recognising the importance of tables in these documents, the proposed \gls{qa} system, leveraging \gls{rag}, incorporates hybrid data and a hierarchical retriever to preserve content integrity and enhance answer extraction.

To conduct an end-to-end evaluation of the \gls{qa} system, the Telco-DPR dataset was released, utilizing a subset of the \gls{3gpp} corpus for assessing the retriever's performance. The methodology of \gls{dhr} demonstrated the best results by effectively searching for context based on the document's hierarchy. This approach achieved an \gls{mrr} of 0.68 and an accuracy of 86.2\%, both at rank 10. Lastly, the \gls{qa} system was compared against a benchmark, demonstrating a 14\% improvement in accuracy on the \gls{mcq} dataset when utilising the more capable \gls{gpt}-4 model, which excels at extracting information from hybrid data.

Future work will develop \gls{qa} systems capable of handling open-ended questions and integrate a multi-modal approach to include visual elements found in technical documents.

\section*{Acknowledgements}
This work was carried out in the scope of the international project 6G Self Organising and Managing Open Radio Access Network (6G-SMART) C2023/2-20, under the CELTIC-NEXT Core Group and the EUREKA Clusters program. The authors would like to thank CELFINET - A Cyient Company and Instituto de Telecomunicações (IT) for their support.

\bibliography{aux/references.bib}

\end{document}